\documentclass[twocolumn]{aastex631}
\usepackage{amsmath}     
\usepackage{wasysym}           
\usepackage{graphicx}
\usepackage{amssymb}
\usepackage[outdir=./]{epstopdf}
\usepackage{mathrsfs}

\usepackage{anyfontsize}
\usepackage{natbib}

\usepackage{natbib}
\usepackage{color}
\usepackage{lipsum}
\usepackage{diagbox}

\shortauthors{Cufari, Coughlin, \& Nixon}
\begin{document}

\title{The Eccentric Nature of Eccentric Tidal Disruption Events}

\author[0000-0001-8429-754X]{M. Cufari}
\affiliation{Department of Physics, Syracuse University, Syracuse, NY 13210, USA}
\email{mcufari@syr.edu}
\author[0000-0003-3765-6401]{Eric R.~Coughlin}
\affiliation{Department of Physics, Syracuse University, Syracuse, NY 13210, USA}
\email{ecoughli@syr.edu}
\author[0000-0002-2137-4146]{C.~J.~Nixon}
\affiliation{Department of Physics and Astronomy, University of Leicester, Leicester, LE1 7RH, UK}

\begin{abstract}
Upon entering the tidal sphere of a supermassive black hole, a star is ripped apart by tides and transformed into a stream of debris. The ultimate fate of that debris, and the properties of the bright flare that is produced and observed, depends on a number of parameters, including the energy of the center of mass of the original star. Here we present the results of a set of smoothed particle hydrodynamics simulations in which a $1~M_\odot $, $\gamma = 5/3$ polytrope is disrupted by a $10^6 ~M_\odot$ supermassive black hole. 
{}{Each simulation has a pericenter distance of $r_{\rm p} = r_{\rm t}$ (i.e., $\beta \equiv r_{\rm t}/r_{\rm p} = 1$ with $r_{\rm t}$ the tidal radius),} and we vary the eccentricity $e$ of the stellar orbit from $e = 0.8$ up to $e = 1.20$ and study the nature of the fallback of debris onto the black hole and the long-term fate of the unbound material.
For simulations with eccentricities $e \lesssim 0.98$, the fallback curve has a distinct, three-peak structure that is induced by self-gravity. For simulations with eccentricities $e \gtrsim 1.06$, the core of the disrupted star reforms following its initial disruption. Our results have implications for, e.g., tidal disruption events produced by supermassive black hole binaries.
\end{abstract}


\section{Introduction} \label{sec:introduction}
When a star enters the tidal radius of a supermassive black hole, defined to be $r_{\rm t} = \left({M_\bullet}/{M_{\star}}\right)^{1/3}R_{\star}$, where $M_{\bullet}$ is the supermassive black hole mass and $M_{\star}$ and $R_{\star}$ are the stellar mass and radius, respectively, the strong tidal forces of the black hole
tear the star apart and transform it into a stream of stellar debris (\citealt{1975Natur.254..295H, 1994ApJ...422..508K}).
Approximately half of the disrupted stellar material is bound to the supermassive black hole in these tidal disruption events (TDEs) and will accrete, producing a luminous, multiwavelength flare \citep{1988Natur.333..523R}. Observations of TDEs over a range of wavelengths and epochs have now been made (e.g., \citealt{bade96}; \citealt{1999A&A...349L..45K}; \citealt{Esquej2007}; \citealp{2009ApJ...698.1367G, 2012Natur.485..217G}; \citealt{Holien2014}; \citealt{2015Natur.526..542M}; \citealt{2015ApJ...798...12V}; \citealt{2016ApJ...819L..25A}; \citealt{2016ApJ...818L..32C}; \citealt{Holoien2016}; \citealt{2016Natur.535..388K}; \citealt{2016Sci...351...62V}; \citealt{2017ApJ...837..153A}; \citealt{2017ApJ...843..106B}; \citealt{2017MNRAS.466.4904B}; \citealt{2017ApJ...851L..47G}; \citealt{2017ApJ...842...29H}; \citealt{2017A&A...598A..29S}; \citealt{2018MNRAS.473.1130B}; \citealt{2018ApJ...856....1P}; \citealt{2019ApJ...873...92B}; \citealt{2019ApJ...879..119H}; \citealt{2019ApJ...883..111H}; \citealt{2019ApJ...887..218L}; \citealt{2019MNRAS.488.1878N}; \citealt{2019Sci...363..531P}; \citealt{2019A&A...630A..98S}; \citealt{2020arXiv201101593H, 2020ApJ...903...31H};\citealt{2020ApJ...898..161H}; \citealt{2020ApJ...889..166J}; \citealt{2020A&A...639A.100K}; \citealt{2020ApJ...891..121L}; \citealt{2021MNRAS.500.1673H}; \citealt{2021ApJ...910..125P}; \citealt{2021ApJ...908....4V}); see also the recent reviews by \cite{2020SSRv..216...81A}; \cite{2020SSRv..216..124V, saxton20}; \cite{2021arXiv210414580G}.

Studies have investigated how the outcome of a TDE depends on the properties of the disrupted star (e.g., \citealt{2012ApJ...757..134M}; \citealt{2013ApJ...767...25G}; \citealt{2019ApJ...882L..26G}; \citealt{2019ApJ...882L..25L}), properties of the black hole (e.g., \citealt{2017MNRAS.465.3840C}; \citealt{2021MNRAS.503.6005W}), and the stellar orbital parameters such as the penetration factor $\beta \equiv r_{\rm p}/r_{\rm t}$ which is the ratio of the pericenter distance, $r_{\rm p}$, to the tidal radius, $r_{\rm t}$ (e.g., \citealt{2013ApJ...767...25G, 2019MNRAS.488.5267D};
{}{\citealt{2020A&A...642A.111C}};
\citealt{2020ApJ...899...36M}; \citealt{2021arXiv210804242N}),  and eccentricity (\citealt{2013MNRAS.434..909H};  \citealt{2016MNRAS.461.3760H}; \citealt{2018ApJ...855..129H};
\citealt{2020A&A...642A.111C};   \citealt{2020ApJ...900....3P}). {}{Stars orbiting a SMBH may fall within the tidal radius of the black hole through two-body interactions from $\sim$ the sphere of influence of the black hole, on the order of $\sim$ pcs; \citealt{1976MNRAS.176..633F}; \citealt{1977ApJ...211..244L}; \citealt{1978ApJ...226.1087C}; \citealt{1999MNRAS.309..447M, stone13}. In this case, stars with trajectories that fall within the tidal radius have low angular momenta, large apocenters relative to their pericenters, and eccentricities that deviate only very slightly from parabolic {}{(i.e., have eccentricities $e\sim 1$)}. However, stars may also diffuse slowly (i.e., over many orbital timescales of the star about the galactic center) in energy space instead of rapidly (i.e., sub-orbital-timescale) in angular momentum space, though the number of such stars is relatively few and is rapidly depleted. In this case, one would expect the energy of the star to be a substantial fraction of the binding energy of the circular orbit at the tidal radius by the time it crosses the tidal sphere, and an eccentricity that satisfies $0 < e < 1$. Finally, a third possibility is that the SMBH that disrupts the star has a binary companion; in this situation, the star may orbit about the binary in a chaotic fashion for many binary orbital periods before passing through the tidal radius of the hole, with its final energy differing from the (presumed zero) energy that it had upon initially encountering the binary. In this situation the energy of the center of mass may be positive or negative upon reaching the tidal radius of the hole \citep{2017MNRAS.465.3840C, coughlin19b} and the eccentricity could be near zero, near parabolic, or hyperbolic.}

Stars on eccentric orbits (i.e., non-parabolic; we include hyperbolic orbits, which have $e > 1$, in the definition of an eccentric orbit) with eccentricities $e$ that satisfy $e < e^{-}_{\rm crit}$ produce no unbound debris, while those with $e > e^{+}_{\rm crit}$ produce no bound debris. The quantity $e_{\rm crit}$ is given by (e.g., {}{\citealt{2018ApJ...855..129H}})

\begin{equation}\label{eq:e_crit}
    e_{\rm crit}^{\pm} = 1 \pm \frac{2}{\beta}\left(\frac{M_{\bullet}}{M_{\star}}\right)^{-1/3}.
\end{equation} 

For stellar orbits with an eccentricity in between these two extremes, the disruption produces both bound and unbound debris, with the ratio of the amount of bound to unbound debris depending on the eccentricity; $e = 1.0$ yields half bound and half unbound (roughly; the binding energy of the original star changes this number slightly; \citealt{1988Natur.333..523R}).
\begin{figure}[t!]
    \centering
    \includegraphics[width=\columnwidth]{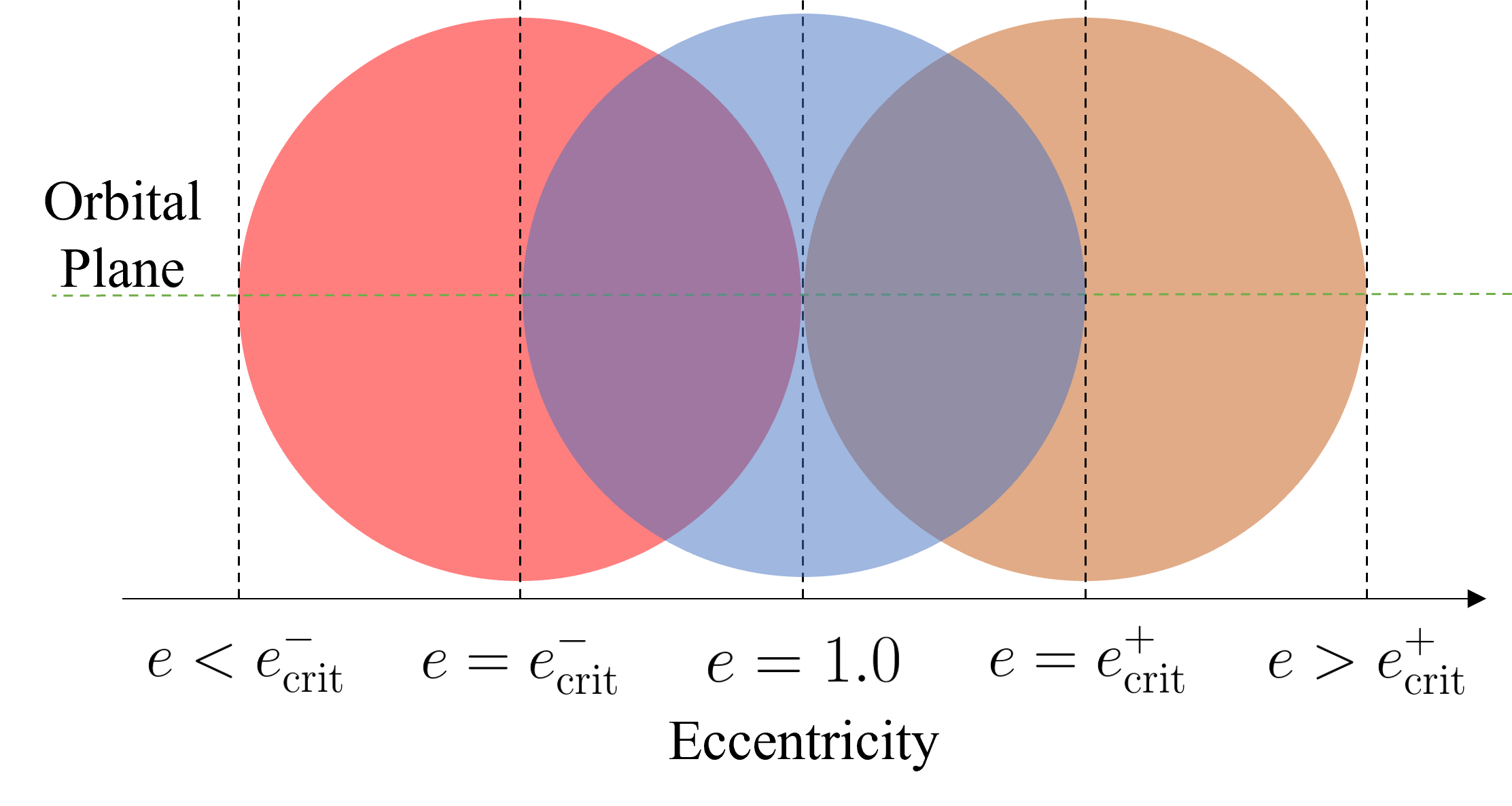}
    \caption{Above is a schematic of a star placed on orbits with different eccentricities. The {}{red (left-most)} star is on an orbit at the lower critical eccentricity $e^{-}_{\rm crit}$, so the outermost edge of the star resides at the zero-energy orbit (i.e., the entire star is just barely bound to the supermassive black hole). The {}{blue (middle)} star is on a parabolic trajectory with the center of mass satisfying $e = 1.0$; roughly half of the material remains bound to the black hole after the star is tidally disrupted for this case. The {}{orange (right-most)} star is on an orbit at the upper critical eccentricity $e^{+}_{\rm crit}$, so the innermost edge is resides at the zero-energy orbit and no material is bound to the supermassive black hole.}
    \label{fig:critical_ecc_stars}
\end{figure}

The complete disruption of a star on a parabolic orbit yields a fallback rate that asymptotically scales as $\propto t^{-5/3}$ at late times ({}{\citealt{1988Natur.333..523R}}\footnote{{}{We thank the referee for drawing our attention to the note in the references of \citet{rees90}, which states, ``A stupid error in this paper [\citealt{1988Natur.333..523R}] led to the late-time decline in the infall rate of the debris as $\propto t^{-5/2}$ rather than $\propto t^{-5/3}$.'' }}; \citealt{ 1989IAUS..136..543P}), where the fallback rate is the rate at which tidally disrupted debris returns to pericenter (and, if circularization and subsequent accretion is efficient, is the rate at which the black hole accretes). For a star on an elliptical orbit, the fallback rate -- even at asymptotically late times -- can be less trivial in its temporal dependence. To investigate this dependence, here we present the results of a set  
of smoothed particle hydrodynamic (SPH) simulations of eccentric TDEs, in which we varied the eccentricity of the initial stellar orbit from highly eccentric $(e = 0.8)$ to highly hyperbolic $(e = 1.20)$. \citet{2020ApJ...900....3P} performed a similar set of simulations and extrapolated their results to determine the fallback rate.
We compare our work to theirs and demonstrate that, because of subtleties related to the self-gravity of the disrupted debris, our method of directly calculating the rate of return of debris to the black hole -- in contrast to their method of predicting the fallback rate from a Gaussian-fitted distribution of debris energies -- captures physical variations in the fallback rate that those authors did not recover. \citet{2020A&A...642A.111C} also considered the disruption of bound stars, taking the energies of their stars to match those of the S-stars in the Galactic Center. Their range of energies was sufficiently distinct from ours that we do not compare our results directly to theirs, other than to say that the qualitative features of their fallback curves are in rough agreement with what we find here.

In Section \ref{sec:analytic} we describe an analytic approach to calculate the mass fallback curves, the frozen-in approximation (as employed in \citealt{2009MNRAS.392..332L}), and apply this approach to eccentric stellar disruptions. In Section \ref{sec:sim_setup} we describe the setup and parameter space of our simulations. In Section \ref{sec:sim_results} we describe the results of the simulations,
and our conclusions and discussion are presented in Section \ref{sec:conclusion}.

\section{Analytic Methods} \label{sec:analytic}
\citet{2009MNRAS.392..332L} showed that the mass fallback rate is dependent on the density distribution of the star in the frozen-in approximation, when the energies of the fluid elements of the star are ``frozen-in" when the stellar center of mass crosses the tidal sphere of the supermassive black hole ({}{note, however, that \citet{2009MNRAS.392..332L} assumed that the energy is frozen in when the stellar center of mass reaches the pericenter distance, which may not necessarily coincide with the tidal radius; see \citealt{lacy82, stone13} for the more physically motivated statement of the frozen-in condition, being that the energy distribution is set at the tidal radius, but see also Figure 2 of \citealt{steinberg19}}).
While we do not rely on this approximation for obtaining our results (see Section \ref{sec:sim_results}), here we briefly consider the implications of the frozen-in model for the case where the stellar orbit is eccentric (and not parabolic, the assumption made in \citealt{2009MNRAS.392..332L}). 

\citet{2009MNRAS.392..332L} showed that the mass fallback rate can be reconstructed by considering the star at the time it passes through the tidal radius and breaking up the stellar material into vertical slices of constant distance from the black hole, which correspond to slices of constant energy (to leading order in the ratio of the tidal radius to the stellar radius, or typically to within $\sim 1\%$). Then the differential amount of mass per unit energy is given by 

\begin{equation}
    \frac{dm}{d\epsilon} = 2\pi \int_\epsilon^1 \rho(\xi)\xi d\xi, \label{eq:dmdeLodato}
\end{equation}
where $\rho$ is the density of the star, expressed in units of average density, $3M_*/4\pi R_*^3$, and $\xi$ is the dimensionless stellar radius as it appears in the Lane-Emden equation \citep{chandraBook}. We follow \citet{2018MNRAS.477.4009D} and let $\tau \equiv t/t_0$ be the dimensionless time since disruption where $t_0$ is given by $t_0 = 2\pi G M_\bullet /(2\Delta\epsilon)^{2/3}$. The dimensionless energy $\epsilon$ is given by 
\begin{equation} \label{eq:epsilon}
    \epsilon = \epsilon_c + \frac{\tau^{-2/3}}{2}
\end{equation}
where $\epsilon_c$ is related to the eccentricity by $$\epsilon_c \equiv \frac{-(1-e)}{(1-e_{\text{crit}}^-)},$$
and 
$$\Delta\epsilon \equiv GM_{\bullet}R_{\star}/r_{\rm t}^2.$$ $\Delta\epsilon$ is the spread in energy imparted by the tidal force across the star in the frozen-in model. By expanding the density as a power series in $\xi$ and integrating equation \eqref{eq:dmdeLodato} one finds
\begin{equation}\label{eq7:dmde_final}
\frac{dm}{d\epsilon} = \sum_{\text{n} \in \text{even}} \frac{\partial^n \rho}{\partial \xi ^n} \frac{1}{(n+2)n!}\left(1 - \left(\epsilon_{c} + \frac{\tau^{-2/3}}{2}\right)^{n+2}\right)
\end{equation}
where the derivative in the density is evaluated at $\xi = 0$ and the sum is taken over all even $n \ge 0$.
In the limiting case that the star is critically bound to the black hole $\epsilon_{c} = -1$. Since $\dot{M} \propto \tau^{-5/3}dm/d\epsilon$ the mass fallback rate in this case follows the power law $\dot{M} \propto \tau^{-7/3}$ at late times -- independent of the density profile of the disrupted star. 

\begin{figure*}[t!]
    \centering
    \includegraphics[width=\textwidth]{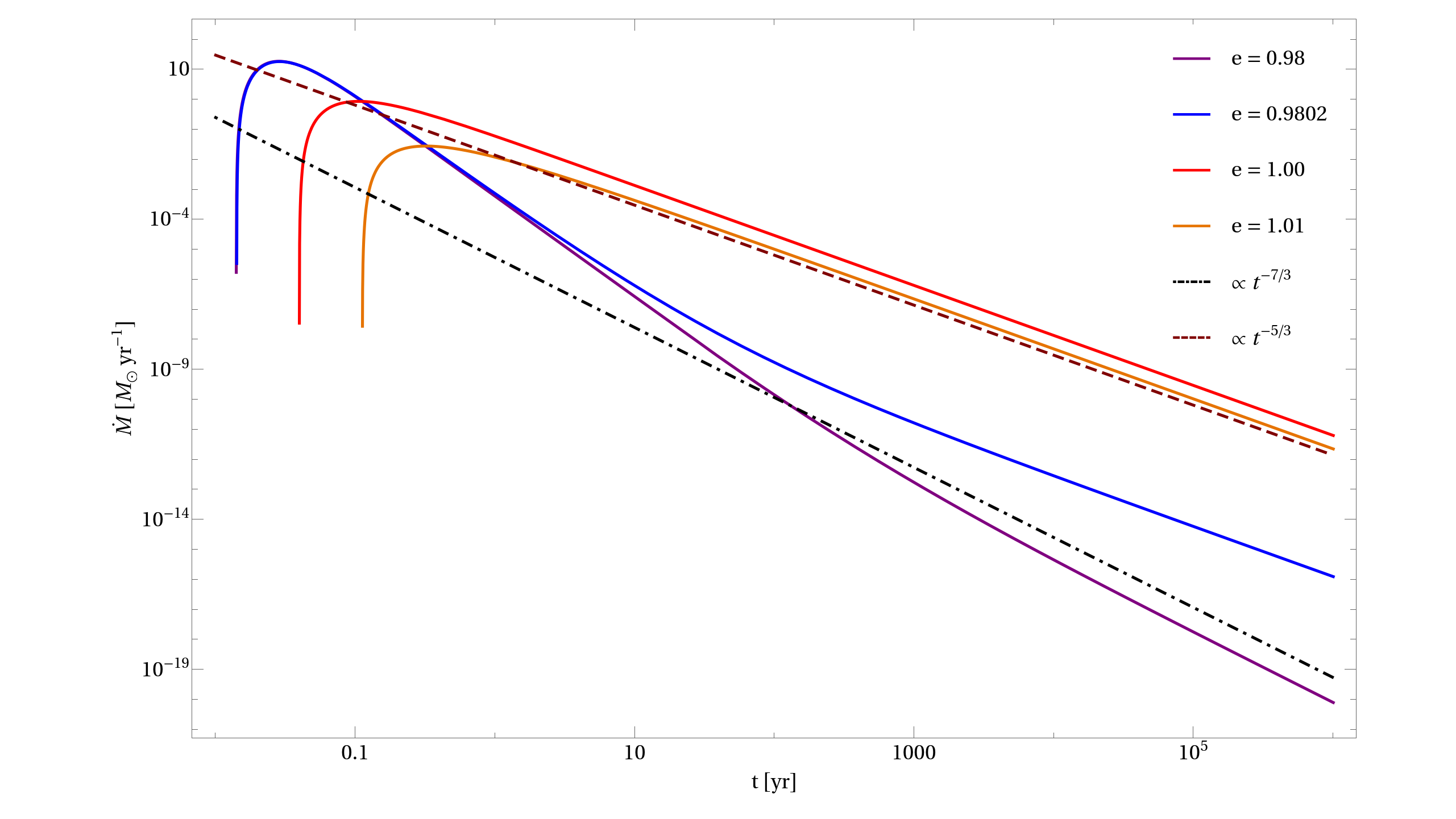}
    \caption{Analytic solutions for the mass fallback rate in the frozen-in approximation. The curves are plotted for a $1 M_\odot$ $\gamma = 5/3$ polytrope disrupted by a $10^6 M_\odot$ black hole on a $\beta = 1.0$ trajectory. {}{Shown in red-brown and black are reference curves for the $t^{-5/3}$ and $ t^{-7/3}$ power laws}. The peak is shifted to higher values, and earlier in time with decreasing eccentricity. The analytic solutions only result in fallback for $e< 1.02$.}
    \label{fig:frozenInAnalytic}
\end{figure*}

In Figure \ref{fig:frozenInAnalytic}, we calculate the fallback curves for a solar-like $\gamma=5/3$ polytrope (i.e., one with a solar mass and radius) placed on different orbits about a black hole of mass $10^6M_{\odot}$ ranging from critically bound, $e = 0.98$ to marginally unbound $e = 1.01$, using this formalism.
For all disruptions the fallback curve is flatter than the $t^{-5/3}$ reference curve at early times as the fallback rate rises and reaches a peak, but the disruption with $e = 1.01$ converges to the canonical power law around $t = $ 10 years. {}{In contrast to the curve corresponding to $e = 1.01$, the mass fallback curve for $e = 1.00$ reaches a  higher peak and converges to the canonical power law at an earlier time. This trend is consistent for disruptions which yield some bound and unbound material.} For $e = 0.9802$ and $e = 0.98$, the fallback curve is much steeper than $t^{-7/3}$ at early times, approaching $t^{-10/3}$ shortly after the peak before flattening out. In the critical case with $e = 0.98$, the fallback rate is proportional to $t^{-7/3}$ at late times, as was shown above.

\begin{figure}[b!]
    \centering
    \includegraphics[width=.495\textwidth]{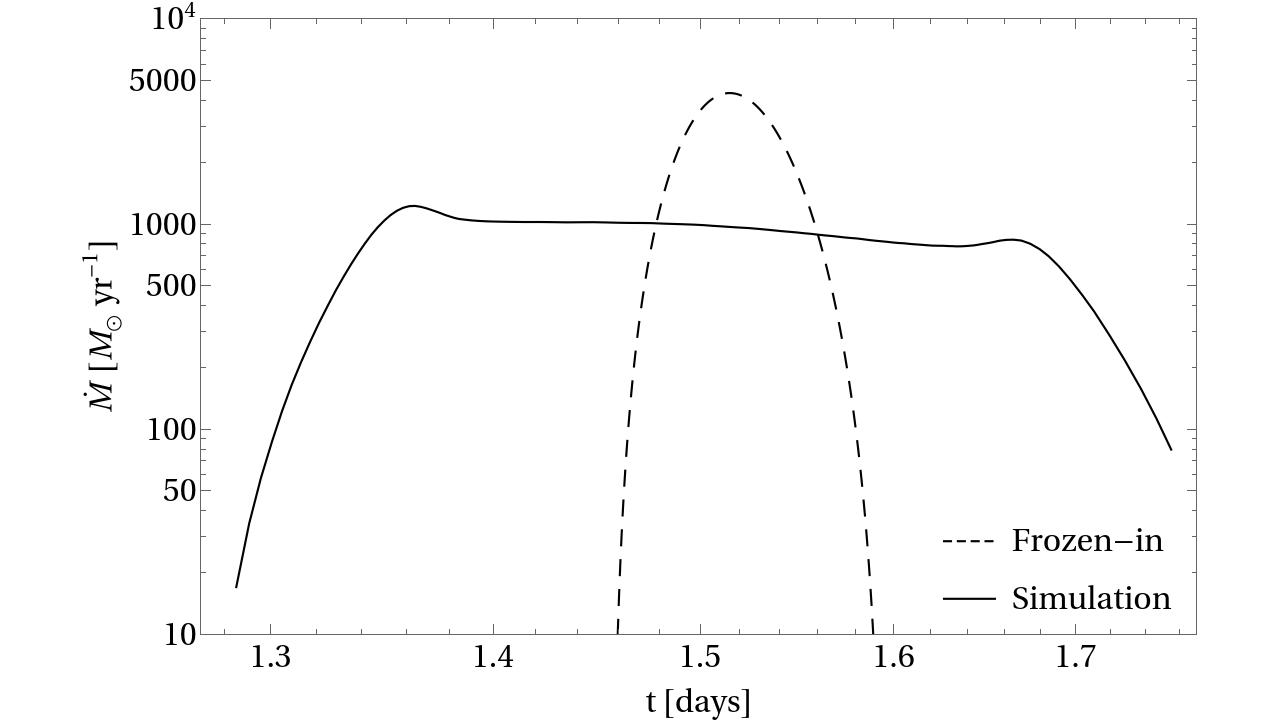}
    \caption{Shown in solid black is the fallback rate for a 1 solar mass $\gamma = 5/3$ polytrope on an $e = 0.80$ orbit about a $10^6\text{M}_\odot$ black hole. The dashed line shows the fallback rate calculated in the frozen-in approximation. The entire fallback episode starts and finishes  within $\sim$13 hours.}
    \label{fig:ecc080_fallback}
\end{figure}
\section{Simulations}
\subsection{Simulation Setup}
\label{sec:sim_setup}
Our simulations are performed with the
SPH code {\sc phantom} \citep{2018PASA...35...31P}. {\sc phantom} has been used in previous studies to simulate TDEs (e.g., \citealt{2015ApJ...808L..11C}; \citealt{2017MNRAS.471L.115C}; \citealt{2018MNRAS.474.3857C};  \citealt{2018MNRAS.478.3016W}; \citealt{2019MNRAS.488.5267D}; \citealt{2019ApJ...872..163G};  \citealt{2019ApJ...882L..26G};  \citealt{2019arXiv191010154L}; \citealt{2020arXiv200804922A}; \citealt{2020MNRAS.495.1374B}; \citealt{2020A&A...642A.111C}; \citealt{2020SSRv..216...63L};  \citealt{2020ApJ...899...36M};     \citealt{2021MNRAS.505L..21T}; \citealt{2021MNRAS.503.6005W}). We use $\sim10^{6}$ particles to model the disruption of a solar-like, $\gamma = 5/3$ polytrope (with $M_{\star} = M_\odot$ and $R_* = R_\odot$) by a $10^{6} M_{\odot}$ supermassive black hole. The details of the numerical method are described in full in \citet{2015ApJ...808L..11C}. {}{Our simulations are performed using an adiabatic equation of state and treat gravity as purely Newtonian. We initialize our simulations by placing the star at a distance of $5r_t$ from the supermassive black hole.} We simulate TDEs with $\beta = 1$ and eccentricities that span $e = 0.9$ to $1.1$ in increments of $0.01$. We also simulate $e = 0.995$, $0.9975$, $1.0025$, and $1.005$ to more finely sample the region around $e = 1.0$ (parabolic), and we also analyze two extreme cases of $e = 0.8$ (highly elliptical) and $e = 1.2$ (highly hyperbolic). All of the particles move with the center of mass initially, which itself has an initial velocity and position appropriate to a Keplerian orbit with the prescribed eccentricity (and pericenter distance of $r_{\rm t}$). For simulations that yield bound material, particles are ``accreted'' once they return within a radius of $150 R_{\rm g} \simeq 3 r_{\rm t}$, where $R_{\rm g} = GM_{\bullet}/c^2$ is the gravitational radius of the hole. In addition, we perform an additional simulation at $e = 1.07$ using $\sim 10^7$ particles in order to check the convergence of our results. We chose this particular simulation to re-simulate at higher resolution for reasons that are detailed below (see Section \ref{subsec:unbound}).

\subsection{Results}
\label{sec:sim_results}

\subsubsection{Bound TDEs} 
\label{subsec:bound}
\begin{figure*}[t!]
    \centering
    \includegraphics[width=\linewidth]{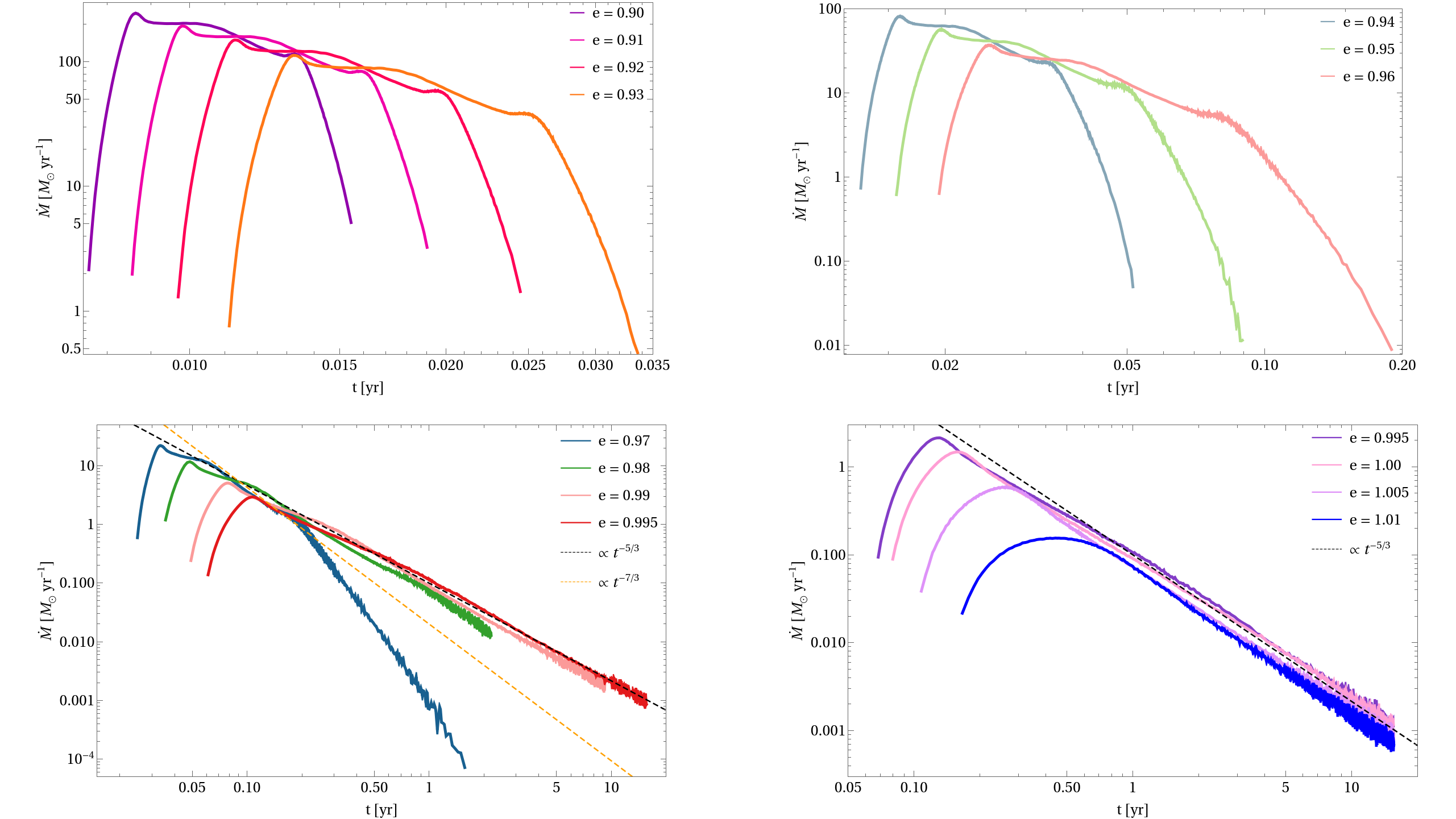}
    \caption{The fallback curves from our simulations, grouped to show the effect of eccentricity on the shape and late time behavior of the fallback. Top-left: The four most bound disruptions (excluding $e= 0.8$ shown in Figure \ref{fig:ecc080_fallback}) at $e = 0.90$, $e = 0.91$, $e = 0.92$ and $e = 0.93$. The secondary elongated peak in each of the four curves corresponds to the accretion of the relatively high-density, central region of the stream, and the entire mass of the star is accreted in a finite time. Top-right: At slightly higher eccentricities the prominent third peak is gradually washed out, as the least bound material to the black hole approaches energies closer to 0. Bottom-left: The transition from {}{the regime where the debris stream is completely bound to the black hole, to the regime where the debris stream is partially bound} occurs between $e = 0.97$ and $e = 0.98$. The primary indicator of this is the late time convergence to the $\propto t^{-5/3}$ power law index. Bottom-right: The second and third peaks are no longer seen in the fallback rate since the material at and beyond the original stellar core is unbound to the black hole. }
    \label{fig:mdot}
\end{figure*}

Figure \ref{fig:ecc080_fallback} shows the fallback rate for the most-bound polytrope in our set of simulations, {}{$e = 0.80$,} where all of the stellar debris has returned to the black hole in a duration of just over 13 hours, while the top-left panel of Figure \ref{fig:mdot} shows the other highly eccentric disruptions we performed. The top-right, bottom-left, and bottom-right panels in Figure \ref{fig:mdot} show the mass fallback curves for the remaining set of simulations, with the eccentricity appropriate to each curve shown in the legends. The most eccentric fallback curves span the order of 3 days before the fallback is complete. The shape of the fallback curves in highly eccentric encounters are composed of a characteristic three-peak structure. These features are due to the action of self-gravity, which causes more material to aggregate near the high-density core of the star and generates distinct ``shoulders'' in the density profile along the stream \citep{2015ApJ...808L..11C}.  The first shoulder in the mass distribution creates the prominent peak in the fallback curve that can be easily seen in all eccentric disruptions, and the second shoulder leads to the peak seen at late times before the fallback rate plummets.  The {}{central peak in the mass distribution creates the secondary peak in the mass fallback rate for disruptions with} $e < 0.99$.

\begin{figure*}[t!]
    \centering
    \includegraphics[width=\linewidth]{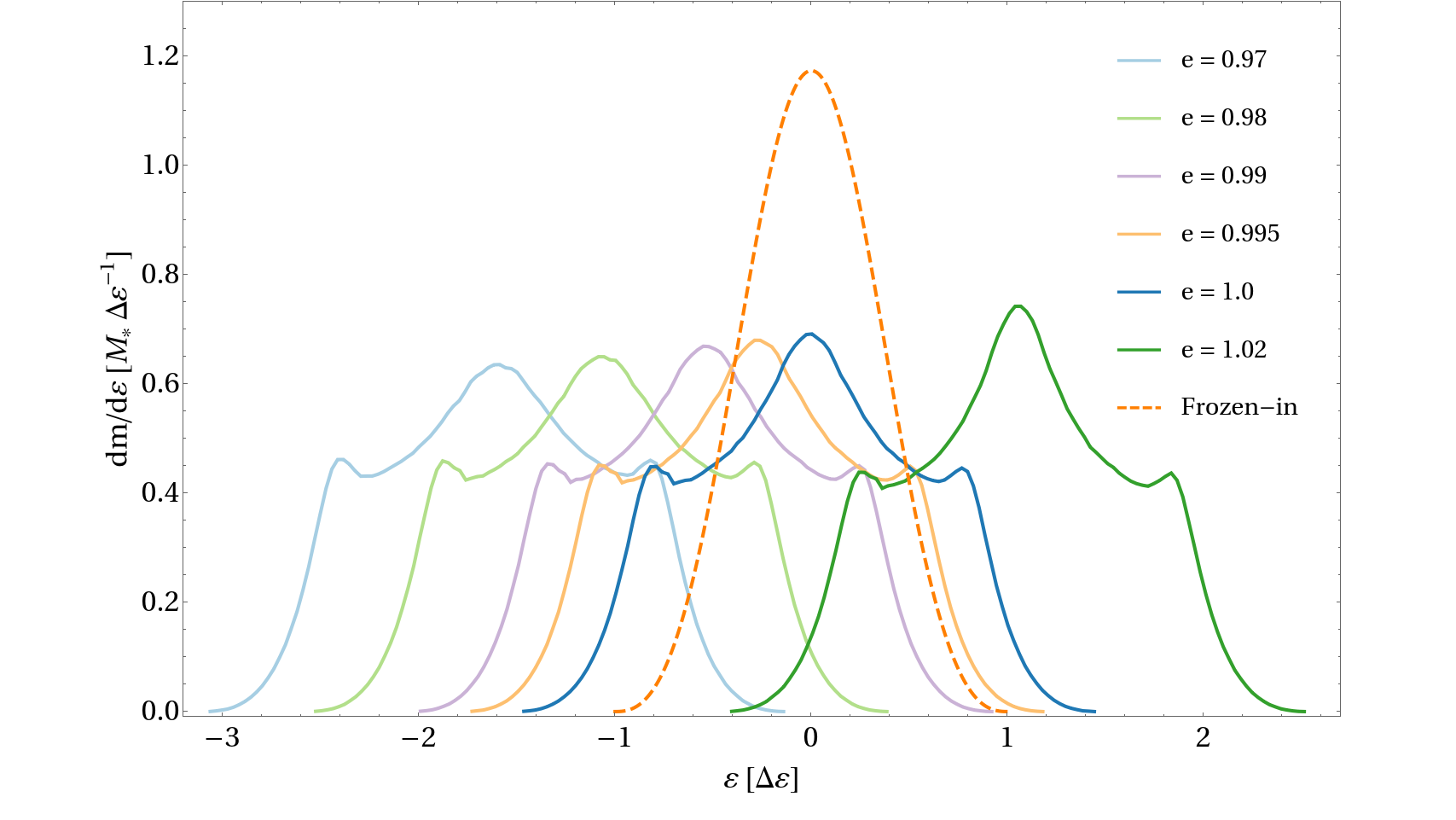}
    \caption{The differential mass per unit energy as a function of Keplerian energy, where the energy is in units of $\Delta \epsilon = GM_{\bullet}R_{\star}/r_{\rm t}^2$, and mass is in units of solar masses. The orange, dashed curve shows the prediction from the frozen-in approximation with $e = 1$ for comparison.}
    \label{fig:dmde}
\end{figure*}

Figure \ref{fig:dmde} shows the mass distribution ${dm}/{d\epsilon}$ for a series of curves  increasing in eccentricity at a fixed time $\sim2.5$ hours after the star reaches pericenter, where $m$ is in units of $M_\odot$ and the energy is in units of $\Delta\epsilon$. 
{}{Under the frozen-in approximation, which ignores hydrodynamical and self-gravitational effects}, the critical eccentricities (given by Equation \ref{eq:e_crit}) are $e_{\rm crit}^- = 0.98$ and $e_{\rm crit}^+ = 1.02$. In our simulations however, we find that the critical eccentricity is $0.97 < e_{\text{crit}} < 0.98$, which likely arises simply from the fact that the energy is not frozen-in precisely at the tidal radius. 
A similar effect occurs at larger eccentricities where the upper critical eccentricity is increased slightly. {}{The dashed curve in this figure shows the frozen-in prediction for $dm/d\epsilon$ for a $\gamma = 5/3$ polytrope and $e = 1$, demonstrating that the energy of the fluid elements is not frozen-in at the tidal radius. Comparing the $e = 1.0$ curve in figure \ref{fig:dmde} to the dashed curve, clearly self-gravitational and hydrodynamical effects have had a non-negligible effect on the mass distribution of the debris stream and hence also the structure of the mass fallback curve.}

\subsubsection{Unbound TDEs}\label{subsec:unbound}
In this section we consider the late time behavior of the debris stream when the entire stream is unbound from the black hole. Our set of simulations included 10 disruptions in this regime, spanning eccentricities from $e = 1.03$ up to $e = 1.20$. We performed 9 simulations at  $10^6$ particles and 1 additional simulation of $e = 1.07$ at $10^7$ particles to verify that the qualitative structure of the stream is unchanged. 

\begin{figure*}[ht!]
    \centering
    \includegraphics[width=\linewidth]{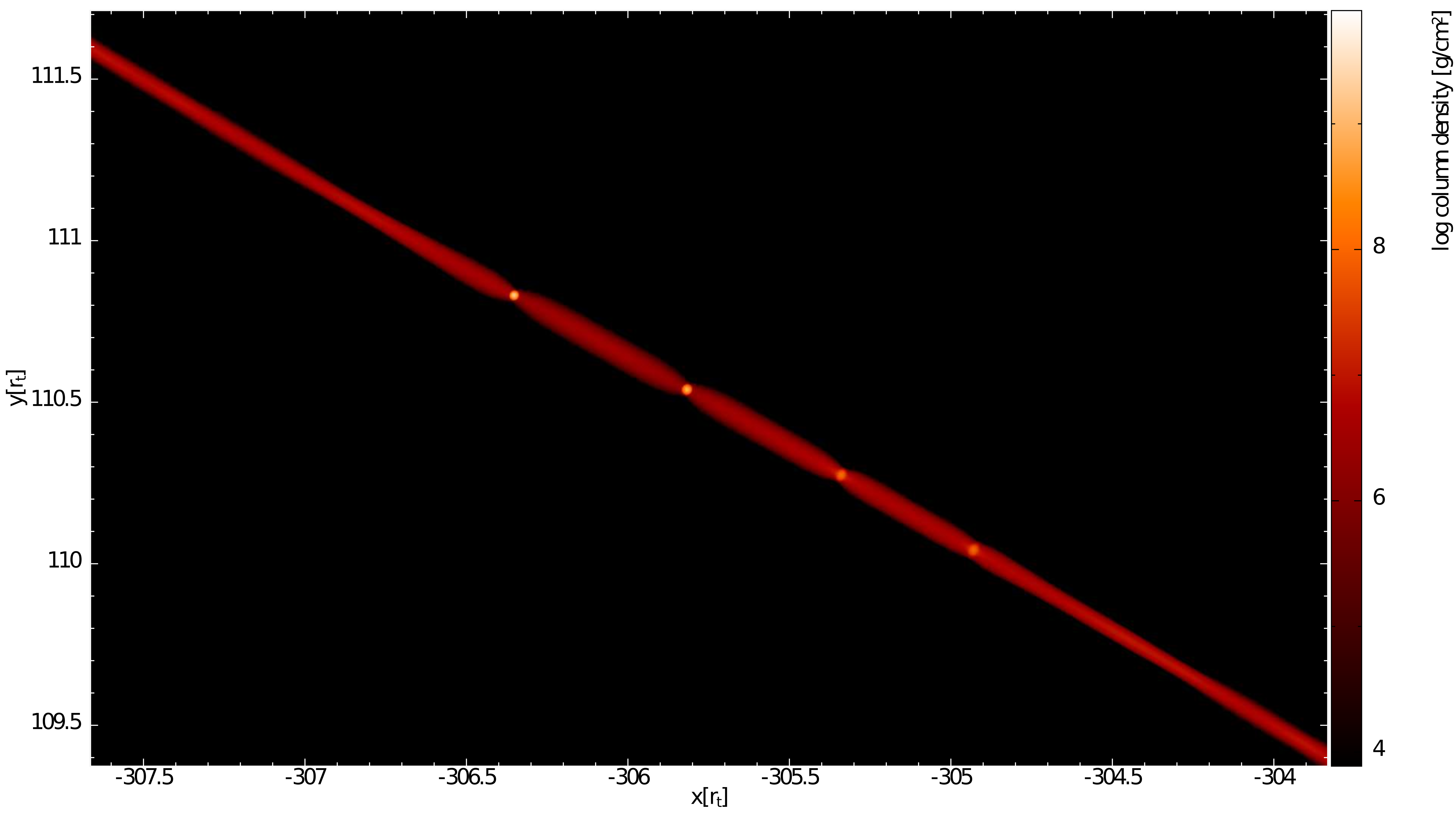}
    \caption{A projection onto the orbital plane of the logarithmic column density, showing the clumps that have formed in the debris stream roughly 50 days since the polytrope was at the distance of closest approach. This disruption took place at $e = 1.06$ and is the largest eccentricity simulation for which multiple cores form; above this eccentricity a single, more massive core reforms.}
    \label{fig:ecc106unstable}
\end{figure*}

Figure \ref{fig:ecc106unstable} shows an image of the column density projected into the orbital plane for an $e = 1.06$ disruption slightly less than 5 days after the polytrope reached pericenter. At this moment in time, four clumps have formed in the debris stream, however, more clumps collapse out of the stream as the simulation progresses. Looking at the trend downward in eccentricity, {}{clumps collapse out of the stream sooner at higher eccentricities, and form clumps of higher density at higher eccentricities. For example, the ratio of the peak density in the debris stream from the $e = 1.06$ to the $e = 1.03$  disruption is $\simeq 550$. By this time 4 clumps have collapsed out of the $e = 1.06$ debris stream and 8 clumps have begun to form, but have not yet completely collapsed out of the $e = 1.05$ stream.} At eccentricities above $e = 1.06$, the star is initially completely destroyed by tides, but at a later time reforms into a single core (or, in the terminology of \citet{2021arXiv210804242N}, a ``zombie core'' rises out of the debris stream).
We verified these results by performing an additional simulation at $10^7$ particles at $e = 1.07$ and found that the structure of the debris stream did not change and a single core reformed at a later time (we discuss the origin of this dichotomy in behavior -- single reformed core versus fragmentation -- in Section \ref{sec:conclusion} below). 
\subsubsection{{}{Zombie core formation and stream fragmentation}}
\label{sec:braaaaains}
{}{The structure of the debris stream in hyperbolic disruptions varies significantly with small changes in eccentricity. In our set of simulations, we found that the stream was unstable and prone to fragmentation. At eccentricities in excess of $e \simeq 1.06$, the core of the disrupted star reforms as a zombie core, i.e., the star was initially completely disrupted but then reformed at a later time \citep{2021arXiv210804242N}; we verified that a single core reforms at $e = 1.07$ when we increased our particle number to $10^7$. At large eccentricities, the disrupted core of the star is able to reform as a result of the more ``glancing" encounter with the tidal sphere of the black hole (loosely, the time spent near the tidal radius, $\sim r_{\rm t}/|v|$, is less as the eccentricity of the stellar orbit increases).} 

{}{As also recently argued in \citet{2021arXiv210804242N}, the dichotomy in the behavior of the stream between eccentricities of $e = 1.06$ and $e = 1.07$, between which the stream goes from fragmentation into a number of small-scale knots to the reformation of a single, massive core, strongly suggests that the underlying mechanism responsible for each of these evolutionary pathways is actually the same, fundamental instability, and in particular the one described in \citet{2020ApJS..247...51C}. The behavior switches between these two outcomes because of the nature of the dispersion relation of the instability, $\sigma(k)$, where $\sigma$ is the growth rate of the instability and $k$ is the wavenumber of the perturbation along the axis of the stream. From Figure 4 of \citet{2020ApJS..247...51C}, $\sigma(k)$ is purely real -- implying instability -- over the range of wavenumbers $0 \le k \le k_{\rm crit}$, where $k_{\rm crit} \sim few$ and is measured in units of the half-width of the stream (its cross-sectional radius), and reaches a maximum at a wavenumber $k_{\rm max} \sim 1$. The stream is therefore unstable to a range of wavenumbers.}

{}{If the stream had equal amounts of power at all wavenumbers for every disruption, we would expect the instability to generate fragments separated roughly by $2\pi H$, as seen in \citet{2020ApJ...896L..38C} for the case of a neutron star black hole merger (and the tidal tail generated therefrom), because it is the most unstable mode that grows fastest and reaches the nonlinear regime -- corresponding to the onset of fragmentation -- earliest; here $H$ is the cross-sectional radius of the stream. However, the peak in the central density of the initial star implies that the resulting power is not uniformly distributed and the distribution of power among wavenumbers varies from disruption to disruption. In particular, when the star is less strongly perturbed near pericenter, which corresponds to when the eccentricity is largest based on our simple timing argument, the width over which the density falls to, e.g., half its value at the center of the stream is smaller, implying that there is more power at larger $k$ when the star is less strongly perturbed. In this case, then, the instability grows faster -- there is more power at the fastest-growing wavenumber -- than when the tidal encounter is deeper and the initial peak in the density is more widely spread out from the tidal shear.} 
{}{The earlier time at which the recollapse occurs implies that the reformed core has a larger mass, and the gravitational force exerted by the reformed core strongly affects the remaining stream around it. For example, the accretion of material onto the core -- and the stronger tidal shear along the stream that is generated by the gravitational field of the core -- stifles further fragmentation, leaving only one core with substantial mass. On the other hand, when the instability occurs at later times, the mass in each of the fragments is less and each one is effectively isolated from the others, allowing many to form before further fragmentation is halted by the increased shear along the stream that is induced by the fragments. It is around the region in parameter space that divides these two outcomes (here the parameter is the eccentricity, but in most studies it is $\beta$) that the final state -- one single core with most of the mass or many isolated fragments with a small fraction of the stream mass -- is less clear and likely dependent on many other variables (e.g., the exact density profile of the star, its initial rotation, the resolution of the simulation). In one case studied in \citet{2021arXiv210804242N}, for example, the stream fragmented into a collection of \emph{five} cores of comparable mass when the $\beta$ of the encounter was around the value that distinguished between zombie-core formation versus global fragmentation (i.e., instead of recollapsing into one massive core or fragmenting into many ($\gg 1$) small-mass ($\ll$ the mass of the original star) fragments, the stream condenses into five knots, each of which has a mass of $\sim 10\%$ of the mass of the original star).}

\subsubsection{Comparison to \citet{2020ApJ...900....3P}}
\label{sec:comps}
As we have already noted, \citet{2020ApJ...900....3P} performed a similar set of hydrodynamical simulations of eccentric TDEs. The simulations of those authors spanned a smaller range in eccentricity, focusing specifically on disruptions that were in between the critical eccentricities, but also varied the point of closest approach (or $\beta$) of the encounter and the type of star (they also modeled $\Gamma = 4/3$ polytropes with $\gamma = 5/3$ for the equation of state).

The main differences between our work and results and those of \citet{2020ApJ...900....3P} are related to differences in the methodologies employed to calculate the fallback rate. In our present work, we evolved the debris stream in time to the point where material began to return to the point of disruption of the original star, after which the gas was artificially accreted by the black hole (artificially from the standpoint that we do not simulate the accretion flow that forms around the black hole and mediates the accretion, but instead we effectively count and bin particles in time as they return in their orbits). Thus we calculate the rate of return of material to the black hole straightforwardly and in a direct, physical manner.

\citet{2020ApJ...900....3P}, on the other hand, calculated the return rate of material to the black hole by using the (widely adopted; e.g., \citealt{evans89}) technique of assuming the energies of the gas parcels comprising the tidally disrupted debris stream are fixed, calculating the $dm/d\epsilon$ curve along the stream, and then using the Keplerian energy-period relationship (cf.~Section \ref{sec:analytic} above) to predict the fallback rate from an earlier time. Furthermore, \citet{2020ApJ...900....3P} did not use their numerically calculated solutions for $dm/d\epsilon$ to extrapolate the fallback rate, but instead used a fitted Gaussian profile for $dm/d\epsilon$ (we speculate that the reason for this was to reduce numerical noise in their solutions for the fallback rate and the power-law index as a function of time, their Figures 8 -- 11{}{, and/or to obtain relatively simple analytic expressions}). 

The assumption that the energy distribution of the gas is fixed (or, indeed, that there exists a well-defined energy that is then relatable to the orbital period of the debris) is not correct when the gas evolves under the influence of a time-dependent force. While this notion was pointed out by \citet{coughlin19} for the specific case of a partial TDE, during which the surviving stellar core exerts a time-dependent, gravitational force on the debris, the same concept applies when the (also time-dependent) self-gravity of the debris plays a dominant role in determining the evolution of the debris (which it does in full disruptions; e.g., \citealt{2015ApJ...808L..11C}). Consequently, the Keplerian energy distribution will change as a function of time, and the orbital period of the debris calculated from that energy will not agree with the true return time of the debris; the time dependence of the energy distribution was made explicit by \citet{norman21}. Nevertheless, provided that the energy distribution is measured at a time well after the debris has exited the tidal sphere of the hole, it is likely that the time dependence of the energy distribution does not generate substantial effects on the measured fallback rate. 

The fact that \citet{2020ApJ...900....3P} used their Gaussian-fitted energy distribution to model their fallback rates likely leads to more substantial differences between our results. For one, the energy distribution fundamentally cannot be a Gaussian because the debris has a finite spread in energy, whereas a Gaussian formally yields infinitely bound and unbound material (in fact, it is unclear why the tightly bound energy tail does not appear at arbitrarily early times in their fallback curves; e.g., their Figure 8). It is also apparent that the behavior of the $dm/d\epsilon$ curve displays much more variation near the center of mass than can be captured by a single Gaussian (i.e., one peak with one prescribed width). Our Figure \ref{fig:dmde}, for example, shows the multiple-peaked structure that arises, we argue, as a consequence of self-gravity, and the imprint that this structure has on the fallback rate -- clearly evident in Figures \ref{fig:ecc080_fallback} and \ref{fig:mdot} --  cannot be captured by a single-Gaussian-fitted energy profile. It is therefore not surprising that the fallback curves of \citet{2020ApJ...900....3P} display a much smoother rise, peak, and decay than ours do (e.g., comparing their Figures 8 and 9 to our Figure \ref{fig:mdot}).
We also point out that \citet{2020ApJ...900....3P} found that their fallback rate power-law indices for partial disruptions were substantially shallower than the recent prediction of \citet{coughlin19}, who used a simple Lagrangian model (making no assumptions about and not relying on any statements concerning energy) to demonstrate that the asymptotic power-law index of any partial TDE is $\simeq -2.26 \simeq -9/4$, effectively independent of the mass of the surviving core (this prediction was recently verified numerically by \citealt{2020ApJ...899...36M} and \citealt{2021arXiv210804242N}). The origin of this discrepancy between their results and this prediction is another manifestation of the incompatibility of the usual, frozen-in approach to determining the fallback rate with the actual physical evolution of a partial TDE: as we pointed out here (see also \citealt{coughlin19}), it is obvious from the expression $dM/dt \propto t^{-5/3}dM/d\epsilon$ that one cannot recover any asymptotic power-law index other than $t^{-5/3}$ because we can expand $dM/d\epsilon$ about the marginally bound radius, i.e., 

\begin{equation}
    \frac{dM}{d\epsilon} = \frac{dM}{d\epsilon}\bigg{|}_{\epsilon = 0}+\frac{d^2M}{d\epsilon^2}\bigg{|}_{\epsilon = 0}\times \epsilon+\mathcal{O}\left(\epsilon^2\right).
\end{equation}
Thus, provided that there is \emph{any mass} (i.e., any non-zero density) at the marginally bound radius within the stream, the solution will asymptote to $\propto t^{-5/3}$ at late times, and only in the situation where $dM/d\epsilon(\epsilon = 0) \equiv 0$, which corresponds to the critically eccentric case, will the falloff be proportional to $t^{-7/3}$ (because $\epsilon \propto t^{-2/3}$; see Figure \ref{fig:frozenInAnalytic} above and our discussion thereof). 

Of course, the frozen-in method (i.e., assuming that $M$ is exclusively a function of $\epsilon$ at any given time) for calculating the fallback rate is fraught with uncertainty when a surviving core (or the stream self-gravity) modifies the dynamical evolution of the tidally disrupted gas, not least because there is no conserved specific energy (by this we mean the energy of a given fluid element; the total energy is conserved, but this conservation does nothing in the way of understanding the physical evolution of the system). As time advances, the density near the marginally bound radius within the stream (which is itself a nebulous concept; see the discussion in \citealt{2013ApJ...767...25G} for their definition of what it means to be bound to the star vs the hole) will decrease more dramatically{}{ than in other regions of the debris stream that are farther from the Hill sphere of the core} owing to the combined effects of the accretion onto the surviving star and the tidal stretching of the stream as induced by the black hole{}{ (and, as a consequence, the $dm/d\epsilon$ curve near the Hill sphere of the core will display a cuspy behavior that reaches a relative minimum; see, e.g., the left-hand panel of Figure 6 of \citealt{gafton19})}. Thus, as the energy distribution is measured at later and later times, the more rapidly declining density in the stream in the vicinity of the star and the concave nature of $dM/d\epsilon$ near the Hill sphere of the surviving core implies that the fallback rate will decline more steeply with time and do so for a longer time before inevitably transitioning to $\propto t^{-5/3}$ (see Figures 9 and 10 of \citealt{2013ApJ...767...25G} for a demonstration of exactly this behavior). The fact that \citet{2020ApJ...900....3P} measured their energy distribution at only $\lesssim 2$ hours post-disruption (in contrast, the return time of the most-bound debris in a $\beta = 1$ disruption under the impulse approximation is $\sim 41$ days, and the numerically simulated return time is found to be closer to $\sim 30$ days) means that the density near the marginally bound radius was still relatively large at that time, and hence their power-law indices ($s(t)$, in their notation) relatively quickly return to the value of $-5/3$ (e.g., their Figure 11). This behavior is, we reiterate, not physical, and is instead an artifact of the method employed to measure the fallback rate.

\section{Summary and Conclusions} \label{sec:conclusion}
We performed 27 simulations of tidal disruption events of $\gamma = 5/3$ polytropes with $\beta = 1$, where we varied the eccecentricity of the encounter from $e = 0.8$ up to $e = 1.2$. At low eccentricities, $e \lesssim 0.98$, the entire mass of the disrupted star {}{returns to the simulated accretion radius, taken to be $\simeq 3r_t$ in all simulations presented here, and all of the particles in the simulation are removed} over a time span of hours to days. However, on trajectories where some material from the star remains unbound to the black hole, the mass fallback rate asymptotically converges to the $t^{-5/3}$ power law. In the limit where the star is critically bound to the black hole, such that all the stellar material is just barely bound to the black hole (the outer edge of the star resides at the $\epsilon = 0$ orbit at the tidal radius in the frozen-in approximation), the mass fallback rate asymptotically approaches $\dot{M} \propto t^{-7/3}$. 

Directly calculating the fallback rates, as we have done here, in eccentric TDEs resolves three peaks in the fallback curve that are apparent when the eccentricity of the disruption is less than the critical eccentricity -- below which all of the mass is bound to the black hole -- which we found numerically to be between $e = 0.97$ and $e = 0.98$. The peaks correspond to the first shoulder, the high-density region around the original stellar core, and second shoulder in the mass distribution, which can be seen in Figure \ref{fig:dmde}. At increasingly higher eccentricities each of these components of the mass distribution becomes unbound from the black hole and hence does not appear in the fallback curve. {}{As the TDE becomes increasingly hyperbolic, the star is less strongly perturbed near pericenter owing to the shorter amount of time spent near the tidal radius, and for eccentricities $e \gtrsim 1.06$, a zombie core reforms after the complete disruption of the star. We also found that the stream is gravitationally unstable and fragments into many, small-mass knots below $e \simeq 1.07$, suggesting that the zombie reformation and the fragmentation are manifestations of the same, underlying instability (see also \citealt{2015ApJ...808L..11C} and \citealt{2021arXiv210804242N}).}

{}{We conclude by noting that, as mentioned in the Introduction (Section \ref{sec:introduction}), the presence of a supermassive black hole binary may produce tidal disruption events in which the disrupted star spans a range of eccentricities, from near-circular, to near-parabolic, to hyperbolic. In this case, the binary companion may play an active role in modifying the fallback dynamics, producing quasi-periodicities in the fallback curve and thereby directly indicating its presence (e.g., \citealt{coughlin19b}). However, another possibility is that the binary separation is sufficiently wide that the only influence of the binary is to modify the initial energy of the disrupted star. When the binary is sufficiently wide that the secondary does not directly modify the fallback dynamics, the detection of a TDE with features characteristic of an eccentric disruption (e.g., the three-peak characteristic structure fallback curves described here) could be a useful diagnostic of the presence of a supermassive black hole binary in the host galaxy.}

\section*{}
{}{We thank the anonymous referee for detailed and useful comments.} M. Cufari~acknowledges this research was supported in part through computational resources provided by Syracuse University and through funding provided by the Syracuse University Office for Undergraduate Research (SOURCE). M. Cufari~also acknowledges assistance and guidance in performing these simulations from Patrick Miles. E.R.C.~acknowledges support from the National Science Foundation through grant AST-2006684. C.J.N acknowledges funding from the European Union’s Horizon 2020 research and innovation program under the Marie Skłodowska-Curie grant agreement No 823823 (Dustbusters RISE project).

\bibliographystyle{aasjournal}
\bibliography{main}

\begin{thebibliography}{}
\expandafter\ifx\csname natexlab\endcsname\relax\def\natexlab#1{#1}\fi
\providecommand{\url}[1]{\href{#1}{#1}}
\providecommand{\dodoi}[1]{doi:~\href{http://doi.org/#1}{\nolinkurl{#1}}}
\providecommand{\doeprint}[1]{\href{http://ascl.net/#1}{\nolinkurl{http://ascl.net/#1}}}
\providecommand{\doarXiv}[1]{\href{https://arxiv.org/abs/#1}{\nolinkurl{https://arxiv.org/abs/#1}}}

\bibitem[{{Alexander} {et~al.}(2016){Alexander}, {Berger}, {Guillochon},
  {Zauderer}, \& {Williams}}]{2016ApJ...819L..25A}
{Alexander}, K.~D., {Berger}, E., {Guillochon}, J., {Zauderer}, B.~A., \&
  {Williams}, P.~K.~G. 2016, \apjl, 819, L25,
  \dodoi{10.3847/2041-8205/819/2/L25}

\bibitem[{{Alexander} {et~al.}(2020){Alexander}, {van Velzen}, {Horesh}, \&
  {Zauderer}}]{2020SSRv..216...81A}
{Alexander}, K.~D., {van Velzen}, S., {Horesh}, A., \& {Zauderer}, B.~A. 2020,
  \ssr, 216, 81, \dodoi{10.1007/s11214-020-00702-w}

\bibitem[{{Alexander} {et~al.}(2017){Alexander}, {Wieringa}, {Berger},
  {Saxton}, \& {Komossa}}]{2017ApJ...837..153A}
{Alexander}, K.~D., {Wieringa}, M.~H., {Berger}, E., {Saxton}, R.~D., \&
  {Komossa}, S. 2017, \apj, 837, 153, \dodoi{10.3847/1538-4357/aa6192}

\bibitem[{{Andalman} {et~al.}(2020){Andalman}, {Liska}, {Tchekhovskoy},
  {Coughlin}, \& {Stone}}]{2020arXiv200804922A}
{Andalman}, Z.~L., {Liska}, M. T.~P., {Tchekhovskoy}, A., {Coughlin}, E.~R., \&
  {Stone}, N. 2020, arXiv e-prints, arXiv:2008.04922.
\newblock \doarXiv{2008.04922}

\bibitem[{{Bade} {et~al.}(1996){Bade}, {Komossa}, \& {Dahlem}}]{bade96}
{Bade}, N., {Komossa}, S., \& {Dahlem}, M. 1996, \aap, 309, L35

\bibitem[{{Blagorodnova} {et~al.}(2019){Blagorodnova}, {Cenko}, {Kulkarni},
  {Arcavi}, {Bloom}, {Duggan}, {Filippenko}, {Fremling}, {Horesh},
  {Hosseinzadeh}, {Karamehmetoglu}, {Levan}, {Masci}, {Nugent}, {Pasham},
  {Veilleux}, {Walters}, {Yan}, \& {Zheng}}]{2019ApJ...873...92B}
{Blagorodnova}, N., {Cenko}, S.~B., {Kulkarni}, S.~R., {et~al.} 2019, \apj,
  873, 92, \dodoi{10.3847/1538-4357/ab04b0}

\bibitem[{{Blanchard} {et~al.}(2017){Blanchard}, {Nicholl}, {Berger},
  {Guillochon}, {Margutti}, {Chornock}, {Alexander}, {Leja}, \&
  {Drout}}]{2017ApJ...843..106B}
{Blanchard}, P.~K., {Nicholl}, M., {Berger}, E., {et~al.} 2017, \apj, 843, 106,
  \dodoi{10.3847/1538-4357/aa77f7}

\bibitem[{{Bonnerot} \& {Lu}(2020)}]{2020MNRAS.495.1374B}
{Bonnerot}, C., \& {Lu}, W. 2020, \mnras, 495, 1374,
  \dodoi{10.1093/mnras/staa1246}

\bibitem[{{Brown} {et~al.}(2017){Brown}, {Holoien}, {Auchettl}, {Stanek},
  {Kochanek}, {Shappee}, {Prieto}, \& {Grupe}}]{2017MNRAS.466.4904B}
{Brown}, J.~S., {Holoien}, T.~W.~S., {Auchettl}, K., {et~al.} 2017, \mnras,
  466, 4904, \dodoi{10.1093/mnras/stx033}

\bibitem[{{Brown} {et~al.}(2018){Brown}, {Kochanek}, {Holoien}, {Stanek},
  {Auchettl}, {Shappee}, {Prieto}, {Morrell}, {Falco}, {Strader}, {Chomiuk},
  {Post}, {Villanueva}, {Mathur}, {Dong}, {Chen}, \&
  {Bose}}]{2018MNRAS.473.1130B}
{Brown}, J.~S., {Kochanek}, C.~S., {Holoien}, T.~W.~S., {et~al.} 2018, \mnras,
  473, 1130, \dodoi{10.1093/mnras/stx2372}

\bibitem[{{Cenko} {et~al.}(2016){Cenko}, {Cucchiara}, {Roth}, {Veilleux},
  {Prochaska}, {Yan}, {Guillochon}, {Maksym}, {Arcavi}, {Butler}, {Filippenko},
  {Fruchter}, {Gezari}, {Kasen}, {Levan}, {Miller}, {Pasham}, {Ramirez-Ruiz},
  {Strubbe}, {Tanvir}, \& {Tombesi}}]{2016ApJ...818L..32C}
{Cenko}, S.~B., {Cucchiara}, A., {Roth}, N., {et~al.} 2016, \apjl, 818, L32,
  \dodoi{10.3847/2041-8205/818/2/L32}

\bibitem[{{Chandrasekhar}(1967)}]{chandraBook}
{Chandrasekhar}, S. 1967, {An introduction to the study of stellar structure}
  (Dover Publications)

\bibitem[{{Clerici} \& {Gomboc}(2020)}]{2020A&A...642A.111C}
{Clerici}, A., \& {Gomboc}, A. 2020, \aap, 642, A111,
  \dodoi{10.1051/0004-6361/202037641}

\bibitem[{{Cohn} \& {Kulsrud}(1978)}]{1978ApJ...226.1087C}
{Cohn}, H., \& {Kulsrud}, R.~M. 1978, \apj, 226, 1087, \dodoi{10.1086/156685}

\bibitem[{{Coughlin} \& {Armitage}(2017)}]{2017MNRAS.471L.115C}
{Coughlin}, E.~R., \& {Armitage}, P.~J. 2017, \mnras, 471, L115,
  \dodoi{10.1093/mnrasl/slx114}

\bibitem[{{Coughlin} \& {Armitage}(2018)}]{2018MNRAS.474.3857C}
---. 2018, \mnras, 474, 3857, \dodoi{10.1093/mnras/stx3039}

\bibitem[{{Coughlin} {et~al.}(2019){Coughlin}, {Armitage}, {Lodato}, \&
  {Nixon}}]{coughlin19b}
{Coughlin}, E.~R., {Armitage}, P.~J., {Lodato}, G., \& {Nixon}, C.~J. 2019,
  \ssr, 215, 45, \dodoi{10.1007/s11214-019-0612-z}

\bibitem[{{Coughlin} {et~al.}(2017){Coughlin}, {Armitage}, {Nixon}, \&
  {Begelman}}]{2017MNRAS.465.3840C}
{Coughlin}, E.~R., {Armitage}, P.~J., {Nixon}, C., \& {Begelman}, M.~C. 2017,
  \mnras, 465, 3840, \dodoi{10.1093/mnras/stw2913}

\bibitem[{{Coughlin} \& {Nixon}(2015)}]{2015ApJ...808L..11C}
{Coughlin}, E.~R., \& {Nixon}, C. 2015, \apjl, 808, L11,
  \dodoi{10.1088/2041-8205/808/1/L11}

\bibitem[{{Coughlin} \& {Nixon}(2019)}]{coughlin19}
{Coughlin}, E.~R., \& {Nixon}, C.~J. 2019, \apjl, 883, L17,
  \dodoi{10.3847/2041-8213/ab412d}

\bibitem[{{Coughlin} \& {Nixon}(2020)}]{2020ApJS..247...51C}
---. 2020, \apjs, 247, 51, \dodoi{10.3847/1538-4365/ab77c2}

\bibitem[{{Coughlin} {et~al.}(2020){Coughlin}, {Nixon}, {Barnes}, {Metzger}, \&
  {Margutti}}]{2020ApJ...896L..38C}
{Coughlin}, E.~R., {Nixon}, C.~J., {Barnes}, J., {Metzger}, B.~D., \&
  {Margutti}, R. 2020, \apjl, 896, L38, \dodoi{10.3847/2041-8213/ab9a4e}

\bibitem[{{Darbha} {et~al.}(2019){Darbha}, {Coughlin}, {Kasen}, \&
  {Nixon}}]{2019MNRAS.488.5267D}
{Darbha}, S., {Coughlin}, E.~R., {Kasen}, D., \& {Nixon}, C. 2019, \mnras, 488,
  5267, \dodoi{10.1093/mnras/stz1923}

\bibitem[{{Darbha} {et~al.}(2018){Darbha}, {Coughlin}, {Kasen}, \&
  {Quataert}}]{2018MNRAS.477.4009D}
{Darbha}, S., {Coughlin}, E.~R., {Kasen}, D., \& {Quataert}, E. 2018, \mnras,
  477, 4009, \dodoi{10.1093/mnras/sty822}

\bibitem[{{Esquej, P.} {et~al.}(2007){Esquej, P.}, {Saxton, R. D.}, {Freyberg,
  M. J.}, {Read, A. M.}, {Altieri, B.}, {Sanchez-Portal, M.}, \& {Hasinger,
  G.}}]{Esquej2007}
{Esquej, P.}, {Saxton, R. D.}, {Freyberg, M. J.}, {et~al.} 2007, A\&A, 462,
  L49, \dodoi{10.1051/0004-6361:20066072}

\bibitem[{{Evans} \& {Kochanek}(1989)}]{evans89}
{Evans}, C.~R., \& {Kochanek}, C.~S. 1989, \apjl, 346, L13,
  \dodoi{10.1086/185567}

\bibitem[{{Frank} \& {Rees}(1976)}]{1976MNRAS.176..633F}
{Frank}, J., \& {Rees}, M.~J. 1976, \mnras, 176, 633,
  \dodoi{10.1093/mnras/176.3.633}

\bibitem[{{Gafton} \& {Rosswog}(2019)}]{gafton19}
{Gafton}, E., \& {Rosswog}, S. 2019, \mnras, 487, 4790,
  \dodoi{10.1093/mnras/stz1530}

\bibitem[{{Gezari}(2021)}]{2021arXiv210414580G}
{Gezari}, S. 2021, arXiv e-prints, arXiv:2104.14580.
\newblock \doarXiv{2104.14580}

\bibitem[{{Gezari} {et~al.}(2017){Gezari}, {Cenko}, \&
  {Arcavi}}]{2017ApJ...851L..47G}
{Gezari}, S., {Cenko}, S.~B., \& {Arcavi}, I. 2017, \apjl, 851, L47,
  \dodoi{10.3847/2041-8213/aaa0c2}

\bibitem[{{Gezari} {et~al.}(2009){Gezari}, {Heckman}, {Cenko}, {Eracleous},
  {Forster}, {Gon{\c{c}}alves}, {Martin}, {Morrissey}, {Neff}, {Seibert},
  {Schiminovich}, \& {Wyder}}]{2009ApJ...698.1367G}
{Gezari}, S., {Heckman}, T., {Cenko}, S.~B., {et~al.} 2009, \apj, 698, 1367,
  \dodoi{10.1088/0004-637X/698/2/1367}

\bibitem[{{Gezari} {et~al.}(2012){Gezari}, {Chornock}, {Rest}, {Huber},
  {Forster}, {Berger}, {Challis}, {Neill}, {Martin}, {Heckman}, {Lawrence},
  {Norman}, {Narayan}, {Foley}, {Marion}, {Scolnic}, {Chomiuk}, {Soderberg},
  {Smith}, {Kirshner}, {Riess}, {Smartt}, {Stubbs}, {Tonry}, {Wood-Vasey},
  {Burgett}, {Chambers}, {Grav}, {Heasley}, {Kaiser}, {Kudritzki}, {Magnier},
  {Morgan}, \& {Price}}]{2012Natur.485..217G}
{Gezari}, S., {Chornock}, R., {Rest}, A., {et~al.} 2012, \nat, 485, 217,
  \dodoi{10.1038/nature10990}

\bibitem[{{Golightly} {et~al.}(2019{\natexlab{a}}){Golightly}, {Coughlin}, \&
  {Nixon}}]{2019ApJ...872..163G}
{Golightly}, E. C.~A., {Coughlin}, E.~R., \& {Nixon}, C.~J. 2019{\natexlab{a}},
  \apj, 872, 163, \dodoi{10.3847/1538-4357/aafd2f}

\bibitem[{{Golightly} {et~al.}(2019{\natexlab{b}}){Golightly}, {Nixon}, \&
  {Coughlin}}]{2019ApJ...882L..26G}
{Golightly}, E.~C.~A., {Nixon}, C.~J., \& {Coughlin}, E.~R. 2019{\natexlab{b}},
  \apjl, 882, L26, \dodoi{10.3847/2041-8213/ab380d}

\bibitem[{{Guillochon} \& {Ramirez-Ruiz}(2013)}]{2013ApJ...767...25G}
{Guillochon}, J., \& {Ramirez-Ruiz}, E. 2013, \apj, 767, 25,
  \dodoi{10.1088/0004-637X/767/1/25}

\bibitem[{{Hayasaki} {et~al.}(2013){Hayasaki}, {Stone}, \&
  {Loeb}}]{2013MNRAS.434..909H}
{Hayasaki}, K., {Stone}, N., \& {Loeb}, A. 2013, \mnras, 434, 909,
  \dodoi{10.1093/mnras/stt871}

\bibitem[{{Hayasaki} {et~al.}(2016){Hayasaki}, {Stone}, \&
  {Loeb}}]{2016MNRAS.461.3760H}
---. 2016, \mnras, 461, 3760, \dodoi{10.1093/mnras/stw1387}

\bibitem[{{Hayasaki} {et~al.}(2018){Hayasaki}, {Zhong}, {Li}, {Berczik}, \&
  {Spurzem}}]{2018ApJ...855..129H}
{Hayasaki}, K., {Zhong}, S., {Li}, S., {Berczik}, P., \& {Spurzem}, R. 2018,
  \apj, 855, 129, \dodoi{10.3847/1538-4357/aab0a5}

\bibitem[{{Hills}(1975)}]{1975Natur.254..295H}
{Hills}, J.~G. 1975, \nat, 254, 295, \dodoi{10.1038/254295a0}

\bibitem[{{Hinkle} {et~al.}(2021){Hinkle}, {Holoien}, {Auchettl}, {Shappee},
  {Neustadt}, {Payne}, {Brown}, {Kochanek}, {Stanek}, {Graham}, {Tucker}, {Do},
  {Anderson}, {Bose}, {Chen}, {Coulter}, {Dimitriadis}, {Dong}, {Foley},
  {Huber}, {Hung}, {Kilpatrick}, {Pignata}, {Piro}, {Rojas-Bravo}, {Siebert},
  {Stalder}, {Thompson}, {Tonry}, {Vallely}, \&
  {Wisniewski}}]{2021MNRAS.500.1673H}
{Hinkle}, J.~T., {Holoien}, T.~W.~S., {Auchettl}, K., {et~al.} 2021, \mnras,
  500, 1673, \dodoi{10.1093/mnras/staa3170}

\bibitem[{Holoien {et~al.}(2014)Holoien, Prieto, Bersier, Kochanek, Stanek,
  Shappee, Grupe, Basu, Beacom, Brimacombe, Brown, Davis, Jencson, Pojmanski,
  \& Szczygieł}]{Holien2014}
Holoien, T. W.-S., Prieto, J.~L., Bersier, D., {et~al.} 2014, Monthly Notices
  of the Royal Astronomical Society, 445, 3263, \dodoi{10.1093/mnras/stu1922}

\bibitem[{Holoien {et~al.}(2016)Holoien, Kochanek, Prieto, Grupe, Chen,
  Godoy-Rivera, Stanek, Shappee, Dong, Brown, Basu, Beacom, Bersier,
  Brimacombe, Carlson, Falco, Johnston, Madore, Pojmanski, \&
  Seibert}]{Holoien2016}
Holoien, T. W.-S., Kochanek, C.~S., Prieto, J.~L., {et~al.} 2016, Monthly
  Notices of the Royal Astronomical Society, 463, 3813,
  \dodoi{10.1093/mnras/stw2272}

\bibitem[{{Holoien} {et~al.}(2019){Holoien}, {Vallely}, {Auchettl}, {Stanek},
  {Kochanek}, {French}, {Prieto}, {Shappee}, {Brown}, {Fausnaugh}, {Dong},
  {Thompson}, {Bose}, {Neustadt}, {Cacella}, {Brimacombe}, {Kendurkar},
  {Beaton}, {Boutsia}, {Chomiuk}, {Connor}, {Morrell}, {Newman}, {Rudie},
  {Shishkovksy}, \& {Strader}}]{2019ApJ...883..111H}
{Holoien}, T. W.~S., {Vallely}, P.~J., {Auchettl}, K., {et~al.} 2019, \apj,
  883, 111, \dodoi{10.3847/1538-4357/ab3c66}

\bibitem[{{Holoien} {et~al.}(2020){Holoien}, {Auchettl}, {Tucker}, {Shappee},
  {Patel}, {Miller-Jones}, {Mockler}, {Groenewald}, {Hinkle}, {Brown},
  {Kochanek}, {Stanek}, {Chen}, {Dong}, {Prieto}, {Thompson}, {Beaton},
  {Connor}, {Cowperthwaite}, {Dahmen}, {French}, {Morrell}, {Buckley},
  {Gromadzki}, {Roy}, {Coulter}, {Dimitriadis}, {Foley}, {Kilpatrick}, {Piro},
  {Rojas-Bravo}, {Siebert}, \& {Velzen}}]{2020ApJ...898..161H}
{Holoien}, T. W.~S., {Auchettl}, K., {Tucker}, M.~A., {et~al.} 2020, \apj, 898,
  161, \dodoi{10.3847/1538-4357/ab9f3d}

\bibitem[{{Hung} {et~al.}(2017){Hung}, {Gezari}, {Blagorodnova}, {Roth},
  {Cenko}, {Kulkarni}, {Horesh}, {Arcavi}, {McCully}, {Yan}, {Lunnan},
  {Fremling}, {Cao}, {Nugent}, \& {Wozniak}}]{2017ApJ...842...29H}
{Hung}, T., {Gezari}, S., {Blagorodnova}, N., {et~al.} 2017, \apj, 842, 29,
  \dodoi{10.3847/1538-4357/aa7337}

\bibitem[{{Hung} {et~al.}(2019){Hung}, {Cenko}, {Roth}, {Gezari}, {Veilleux},
  {van Velzen}, {Gaskell}, {Foley}, {Blagorodnova}, {Yan}, {Graham}, {Brown},
  {Siebert}, {Frederick}, {Ward}, {Gatkine}, {Gal-Yam}, {Yang}, {Schulze},
  {Dimitriadis}, {Kupfer}, {Shupe}, {Rusholme}, {Masci}, {Riddle}, {Soumagnac},
  {van Roestel}, \& {Dekany}}]{2019ApJ...879..119H}
{Hung}, T., {Cenko}, S.~B., {Roth}, N., {et~al.} 2019, \apj, 879, 119,
  \dodoi{10.3847/1538-4357/ab24de}

\bibitem[{{Hung} {et~al.}(2020{\natexlab{a}}){Hung}, {Foley}, {Veilleux},
  {Cenko}, {Dai}, {Auchettl}, {Brink}, {Dimitriadis}, {Filippenko}, {Gezari},
  {Holoien}, {Kilpatrick}, {Mockler}, {Piro}, {Ramirez-Ruiz}, {Rojas-Bravo},
  {Siebert}, {van Velzen}, \& {Zheng}}]{2020arXiv201101593H}
{Hung}, T., {Foley}, R.~J., {Veilleux}, S., {et~al.} 2020{\natexlab{a}}, arXiv
  e-prints, arXiv:2011.01593.
\newblock \doarXiv{2011.01593}

\bibitem[{{Hung} {et~al.}(2020{\natexlab{b}}){Hung}, {Foley}, {Ramirez-Ruiz},
  {Dai}, {Auchettl}, {Kilpatrick}, {Mockler}, {Brown}, {Coulter},
  {Dimitriadis}, {Holoien}, {Law-Smith}, {Piro}, {Rest}, {Rojas-Bravo}, \&
  {Siebert}}]{2020ApJ...903...31H}
{Hung}, T., {Foley}, R.~J., {Ramirez-Ruiz}, E., {et~al.} 2020{\natexlab{b}},
  \apj, 903, 31, \dodoi{10.3847/1538-4357/abb606}

\bibitem[{{Jonker} {et~al.}(2020){Jonker}, {Stone}, {Generozov}, {van Velzen},
  \& {Metzger}}]{2020ApJ...889..166J}
{Jonker}, P.~G., {Stone}, N.~C., {Generozov}, A., {van Velzen}, S., \&
  {Metzger}, B. 2020, \apj, 889, 166, \dodoi{10.3847/1538-4357/ab659c}

\bibitem[{{Kajava} {et~al.}(2020){Kajava}, {Giustini}, {Saxton}, \&
  {Miniutti}}]{2020A&A...639A.100K}
{Kajava}, J. J.~E., {Giustini}, M., {Saxton}, R.~D., \& {Miniutti}, G. 2020,
  \aap, 639, A100, \dodoi{10.1051/0004-6361/202038165}

\bibitem[{{Kara} {et~al.}(2016){Kara}, {Miller}, {Reynolds}, \&
  {Dai}}]{2016Natur.535..388K}
{Kara}, E., {Miller}, J.~M., {Reynolds}, C., \& {Dai}, L. 2016, \nat, 535, 388,
  \dodoi{10.1038/nature18007}

\bibitem[{{Kochanek}(1994)}]{1994ApJ...422..508K}
{Kochanek}, C.~S. 1994, \apj, 422, 508, \dodoi{10.1086/173745}

\bibitem[{{Komossa} \& {Greiner}(1999)}]{1999A&A...349L..45K}
{Komossa}, S., \& {Greiner}, J. 1999, \aap, 349, L45.
\newblock \doarXiv{astro-ph/9908216}

\bibitem[{{Lacy} {et~al.}(1982){Lacy}, {Townes}, \& {Hollenbach}}]{lacy82}
{Lacy}, J.~H., {Townes}, C.~H., \& {Hollenbach}, D.~J. 1982, \apj, 262, 120,
  \dodoi{10.1086/160402}

\bibitem[{{Law-Smith} {et~al.}(2019){Law-Smith}, {Guillochon}, \&
  {Ramirez-Ruiz}}]{2019ApJ...882L..25L}
{Law-Smith}, J., {Guillochon}, J., \& {Ramirez-Ruiz}, E. 2019, \apjl, 882, L25,
  \dodoi{10.3847/2041-8213/ab379a}

\bibitem[{{Leloudas} {et~al.}(2019){Leloudas}, {Dai}, {Arcavi}, {Vreeswijk},
  {Mockler}, {Roy}, {Malesani}, {Schulze}, {Wevers}, {Fraser}, {Ramirez-Ruiz},
  {Auchettl}, {Burke}, {Cannizzaro}, {Charalampopoulos}, {Chen}, {Cikota},
  {Della Valle}, {Galbany}, {Gromadzki}, {Heintz}, {Hiramatsu}, {Jonker},
  {Kostrzewa-Rutkowska}, {Maguire}, {Mandel}, {Nicholl}, {Onori}, {Roth},
  {Smartt}, {Wyrzykowski}, \& {Young}}]{2019ApJ...887..218L}
{Leloudas}, G., {Dai}, L., {Arcavi}, I., {et~al.} 2019, \apj, 887, 218,
  \dodoi{10.3847/1538-4357/ab5792}

\bibitem[{{Li} {et~al.}(2020){Li}, {Saxton}, {Yuan}, {Sun}, {Liu}, {Jiang},
  {Cheng}, {Zhou}, {Komossa}, \& {Jin}}]{2020ApJ...891..121L}
{Li}, D., {Saxton}, R.~D., {Yuan}, W., {et~al.} 2020, \apj, 891, 121,
  \dodoi{10.3847/1538-4357/ab744a}

\bibitem[{{Lightman} \& {Shapiro}(1977)}]{1977ApJ...211..244L}
{Lightman}, A.~P., \& {Shapiro}, S.~L. 1977, \apj, 211, 244,
  \dodoi{10.1086/154925}

\bibitem[{{Liptai} {et~al.}(2019){Liptai}, {Price}, {Mandel}, \&
  {Lodato}}]{2019arXiv191010154L}
{Liptai}, D., {Price}, D.~J., {Mandel}, I., \& {Lodato}, G. 2019, arXiv
  e-prints, arXiv:1910.10154.
\newblock \doarXiv{1910.10154}

\bibitem[{{Lodato} {et~al.}(2020){Lodato}, {Cheng}, {Bonnerot}, \&
  {Dai}}]{2020SSRv..216...63L}
{Lodato}, G., {Cheng}, R.~M., {Bonnerot}, C., \& {Dai}, J.~L. 2020, \ssr, 216,
  63, \dodoi{10.1007/s11214-020-00697-4}

\bibitem[{{Lodato} {et~al.}(2009){Lodato}, {King}, \&
  {Pringle}}]{2009MNRAS.392..332L}
{Lodato}, G., {King}, A.~R., \& {Pringle}, J.~E. 2009, \mnras, 392, 332,
  \dodoi{10.1111/j.1365-2966.2008.14049.x}

\bibitem[{{MacLeod} {et~al.}(2012){MacLeod}, {Guillochon}, \&
  {Ramirez-Ruiz}}]{2012ApJ...757..134M}
{MacLeod}, M., {Guillochon}, J., \& {Ramirez-Ruiz}, E. 2012, \apj, 757, 134,
  \dodoi{10.1088/0004-637X/757/2/134}

\bibitem[{{Magorrian} \& {Tremaine}(1999)}]{1999MNRAS.309..447M}
{Magorrian}, J., \& {Tremaine}, S. 1999, \mnras, 309, 447,
  \dodoi{10.1046/j.1365-8711.1999.02853.x}

\bibitem[{{Miles} {et~al.}(2020){Miles}, {Coughlin}, \&
  {Nixon}}]{2020ApJ...899...36M}
{Miles}, P.~R., {Coughlin}, E.~R., \& {Nixon}, C.~J. 2020, \apj, 899, 36,
  \dodoi{10.3847/1538-4357/ab9c9f}

\bibitem[{{Miller} {et~al.}(2015){Miller}, {Kaastra}, {Miller}, {Reynolds},
  {Brown}, {Cenko}, {Drake}, {Gezari}, {Guillochon}, {Gultekin}, {Irwin},
  {Levan}, {Maitra}, {Maksym}, {Mushotzky}, {O'Brien}, {Paerels}, {de Plaa},
  {Ramirez-Ruiz}, {Strohmayer}, \& {Tanvir}}]{2015Natur.526..542M}
{Miller}, J.~M., {Kaastra}, J.~S., {Miller}, M.~C., {et~al.} 2015, \nat, 526,
  542, \dodoi{10.1038/nature15708}

\bibitem[{{Nicholl} {et~al.}(2019){Nicholl}, {Blanchard}, {Berger}, {Gomez},
  {Margutti}, {Alexander}, {Guillochon}, {Leja}, {Chornock}, {Snios},
  {Auchettl}, {Bruce}, {Challis}, {D'Orazio}, {Drout}, {Eftekhari}, {Foley},
  {Graur}, {Kilpatrick}, {Lawrence}, {Piro}, {Rojas-Bravo}, {Ross}, {Short},
  {Smartt}, {Smith}, \& {Stalder}}]{2019MNRAS.488.1878N}
{Nicholl}, M., {Blanchard}, P.~K., {Berger}, E., {et~al.} 2019, \mnras, 488,
  1878, \dodoi{10.1093/mnras/stz1837}

\bibitem[{{Nixon} {et~al.}(2021){Nixon}, {Coughlin}, \&
  {Miles}}]{2021arXiv210804242N}
{Nixon}, C., {Coughlin}, E., \& {Miles}, P. 2021, arXiv e-prints,
  arXiv:2108.04242.
\newblock \doarXiv{2108.04242}

\bibitem[{{Norman} {et~al.}(2021){Norman}, {Nixon}, \& {Coughlin}}]{norman21}
{Norman}, S. M.~J., {Nixon}, C.~J., \& {Coughlin}, E.~R. 2021, \apj, Under
  Review

\bibitem[{{Park} \& {Hayasaki}(2020)}]{2020ApJ...900....3P}
{Park}, G., \& {Hayasaki}, K. 2020, \apj, 900, 3,
  \dodoi{10.3847/1538-4357/ab9ebb}

\bibitem[{{Pasham} \& {van Velzen}(2018)}]{2018ApJ...856....1P}
{Pasham}, D.~R., \& {van Velzen}, S. 2018, \apj, 856, 1,
  \dodoi{10.3847/1538-4357/aab361}

\bibitem[{{Pasham} {et~al.}(2019){Pasham}, {Remillard}, {Fragile}, {Franchini},
  {Stone}, {Lodato}, {Homan}, {Chakrabarty}, {Baganoff}, {Steiner}, {Coughlin},
  \& {Pasham}}]{2019Sci...363..531P}
{Pasham}, D.~R., {Remillard}, R.~A., {Fragile}, P.~C., {et~al.} 2019, Science,
  363, 531, \dodoi{10.1126/science.aar7480}

\bibitem[{{Payne} {et~al.}(2021){Payne}, {Shappee}, {Hinkle}, {Vallely},
  {Kochanek}, {Holoien}, {Auchettl}, {Stanek}, {Thompson}, {Neustadt},
  {Tucker}, {Armstrong}, {Brimacombe}, {Cacella}, {Cornect}, {Denneau},
  {Fausnaugh}, {Flewelling}, {Grupe}, {Heinze}, {Lopez}, {Monard}, {Prieto},
  {Schneider}, {Sheppard}, {Tonry}, \& {Weiland}}]{2021ApJ...910..125P}
{Payne}, A.~V., {Shappee}, B.~J., {Hinkle}, J.~T., {et~al.} 2021, \apj, 910,
  125, \dodoi{10.3847/1538-4357/abe38d}

\bibitem[{{Phinney}(1989)}]{1989IAUS..136..543P}
{Phinney}, E.~S. 1989, in The Center of the Galaxy, ed. M.~{Morris}, Vol. 136,
  543

\bibitem[{{Price} {et~al.}(2018){Price}, {Wurster}, {Tricco}, {Nixon},
  {Toupin}, {Pettitt}, {Chan}, {Mentiplay}, {Laibe}, {Glover}, {Dobbs},
  {Nealon}, {Liptai}, {Worpel}, {Bonnerot}, {Dipierro}, {Ballabio}, {Ragusa},
  {Federrath}, {Iaconi}, {Reichardt}, {Forgan}, {Hutchison}, {Constantino},
  {Ayliffe}, {Hirsh}, \& {Lodato}}]{2018PASA...35...31P}
{Price}, D.~J., {Wurster}, J., {Tricco}, T.~S., {et~al.} 2018, \pasa, 35, e031,
  \dodoi{10.1017/pasa.2018.25}

\bibitem[{{Rees}(1988)}]{1988Natur.333..523R}
{Rees}, M.~J. 1988, \nat, 333, 523, \dodoi{10.1038/333523a0}

\bibitem[{{Rees}(1990)}]{rees90}
---. 1990, Science, 247, 817, \dodoi{10.1126/science.247.4944.817}

\bibitem[{{Saxton} {et~al.}(2020){Saxton}, {Komossa}, {Auchettl}, \&
  {Jonker}}]{saxton20}
{Saxton}, R., {Komossa}, S., {Auchettl}, K., \& {Jonker}, P.~G. 2020, \ssr,
  216, 85, \dodoi{10.1007/s11214-020-00708-4}

\bibitem[{{Saxton} {et~al.}(2017){Saxton}, {Read}, {Komossa}, {Lira},
  {Alexander}, \& {Wieringa}}]{2017A&A...598A..29S}
{Saxton}, R.~D., {Read}, A.~M., {Komossa}, S., {et~al.} 2017, \aap, 598, A29,
  \dodoi{10.1051/0004-6361/201629015}

\bibitem[{{Saxton} {et~al.}(2019){Saxton}, {Read}, {Komossa}, {Lira},
  {Alexander}, {Steele}, {Oca{\~n}a}, {Berger}, \&
  {Blanchard}}]{2019A&A...630A..98S}
---. 2019, \aap, 630, A98, \dodoi{10.1051/0004-6361/201935650}

\bibitem[{{Steinberg} {et~al.}(2019){Steinberg}, {Coughlin}, {Stone}, \&
  {Metzger}}]{steinberg19}
{Steinberg}, E., {Coughlin}, E.~R., {Stone}, N.~C., \& {Metzger}, B.~D. 2019,
  \mnras, 485, L146, \dodoi{10.1093/mnrasl/slz048}

\bibitem[{{Stone} {et~al.}(2013){Stone}, {Sari}, \& {Loeb}}]{stone13}
{Stone}, N., {Sari}, R., \& {Loeb}, A. 2013, \mnras, 435, 1809,
  \dodoi{10.1093/mnras/stt1270}

\bibitem[{{Trevascus} {et~al.}(2021){Trevascus}, {Price}, {Nealon}, {Liptai},
  {Manser}, \& {Veras}}]{2021MNRAS.505L..21T}
{Trevascus}, D., {Price}, D.~J., {Nealon}, R., {et~al.} 2021, \mnras, 505, L21,
  \dodoi{10.1093/mnrasl/slab043}

\bibitem[{{van Velzen} {et~al.}(2020){van Velzen}, {Holoien}, {Onori}, {Hung},
  \& {Arcavi}}]{2020SSRv..216..124V}
{van Velzen}, S., {Holoien}, T. W.~S., {Onori}, F., {Hung}, T., \& {Arcavi}, I.
  2020, \ssr, 216, 124, \dodoi{10.1007/s11214-020-00753-z}

\bibitem[{{van Velzen} {et~al.}(2016){van Velzen}, {Anderson}, {Stone},
  {Fraser}, {Wevers}, {Metzger}, {Jonker}, {van der Horst}, {Staley}, {Mendez},
  {Miller-Jones}, {Hodgkin}, {Campbell}, \& {Fender}}]{2016Sci...351...62V}
{van Velzen}, S., {Anderson}, G.~E., {Stone}, N.~C., {et~al.} 2016, Science,
  351, 62, \dodoi{10.1126/science.aad1182}

\bibitem[{{van Velzen} {et~al.}(2021){van Velzen}, {Gezari}, {Hammerstein},
  {Roth}, {Frederick}, {Ward}, {Hung}, {Cenko}, {Stein}, {Perley}, {Taggart},
  {Foley}, {Sollerman}, {Blagorodnova}, {Andreoni}, {Bellm}, {Brinnel}, {De},
  {Dekany}, {Feeney}, {Fremling}, {Giomi}, {Golkhou}, {Graham}, {Ho},
  {Kasliwal}, {Kilpatrick}, {Kulkarni}, {Kupfer}, {Laher}, {Mahabal}, {Masci},
  {Miller}, {Nordin}, {Riddle}, {Rusholme}, {van Santen}, {Sharma}, {Shupe}, \&
  {Soumagnac}}]{2021ApJ...908....4V}
{van Velzen}, S., {Gezari}, S., {Hammerstein}, E., {et~al.} 2021, \apj, 908, 4,
  \dodoi{10.3847/1538-4357/abc258}

\bibitem[{{Vink{\'o}} {et~al.}(2015){Vink{\'o}}, {Yuan}, {Quimby}, {Wheeler},
  {Ramirez-Ruiz}, {Guillochon}, {Chatzopoulos}, {Marion}, \&
  {Akerlof}}]{2015ApJ...798...12V}
{Vink{\'o}}, J., {Yuan}, F., {Quimby}, R.~M., {et~al.} 2015, \apj, 798, 12,
  \dodoi{10.1088/0004-637X/798/1/12}

\bibitem[{{Wang} {et~al.}(2021){Wang}, {Perna}, \&
  {Armitage}}]{2021MNRAS.503.6005W}
{Wang}, Y.-H., {Perna}, R., \& {Armitage}, P.~J. 2021, \mnras, 503, 6005,
  \dodoi{10.1093/mnras/stab802}

\bibitem[{{Wu} {et~al.}(2018){Wu}, {Coughlin}, \&
  {Nixon}}]{2018MNRAS.478.3016W}
{Wu}, S., {Coughlin}, E.~R., \& {Nixon}, C. 2018, \mnras, 478, 3016,
  \dodoi{10.1093/mnras/sty971}

\end{thebibliography}

\end{document}